\documentclass[proceedings]{JHEP}
\conference{Corfu Summer Institute on Elementary Particle Physics, 1998}  
\title{
Dynamical Symmetry Breaking with Mirror Fermions}
\author{ George Triantaphyllou
$\;$\\~  \\{\it Institut f\"ur Theoretische Physik, Technische 
Universit\"at M\"unchen}\\
{\it James-Franck-Strasse, D-85748 Garching, GERMANY } \\
E-mail: \email{georg@ph.tum.de}} 
\abstract
{
Mirror fermions  
with masses around the weak scale 
could break dynamically the 
electro-weak symmetry if they were coupled with a new strong interaction.
The purpose of this talk is to show what sort of dynamics are
needed in order to render such theories 
phenomenologically viable. 
} 
\begin{document}
\section{Introduction}

A first speculation for the existence of mirror fermions appeared
in the classical paper on parity violation \cite{Lee} that led to 
the V-A interaction models. Efforts to 
eliminate completely  mirror fermions from nature are for some reminiscent 
of efforts several decades ago to identify the anti-electron with
the proton, and amounts to not realising that particles consistent with  
natural symmetries could actually exist independently.
Such a gauge group and fermion extension,
apart from fitting nicely into unification schemata, 
restores in a certain sense
the left-right symmetry missing in the \newline standard-model
or in the simplest left-right symmetric models. 
In this way it also
provides a well-defined continuum limit of the theory, something 
which is usually problematic due to the Nielsen-Ninomiya theorem 
\cite{Nielsen}. 

The left-right symmetric approach to standard-model 
extensions renders the baryon-lepton number 
symmetry $U(1)_{B-L}$
more natural by gauging it, and has also been proposed
as a solution to the strong CP problem  when 
accompanied with the introduction of 
mirror fermions \cite{Barr}. Moreover, it proves
to be economical by identifying the source of the strong dynamics which 
break the electro-weak symmetry dynamically with
a ``horizontal" generation gauge group in the mirror sector
which, apart from preventing the pairing-up of the standard-model
generations with the mirror ones, provides also the
intra-generation mass hierarchies. 

Furthermore in 
superstring-inspired unification, possibly 
connected to $N=2$ supergravity, the standard-model fermions  
have both mirror
and supersymmetric partners. The present approach corresponds to 
breaking  supersymmetry and leaving the
supersymmetric partners  close to the unification scale, and
bringing the mirror partners down to the weak scale, altering 
thus radically the expected phenomenology. 
 
In this talk, new dynamics needed to make such a
mechanism phenomenologically viable are discussed. 
In particular, it  proves  
necessary to review somewhat 
the dynamical assumptions made in Ref.\cite{litri}.  
In that work it was unclear why the characteristic scale
of the strong group responsible for the fermion gauge-invariant masses
happened to be so close to the scale where 
the strong interactions breaking electro-weak symmetry 
became critical. Furthermore, the
previous model 
could not provide a see-saw mechanism for the standard-model neutrinos,
coupling unification would be difficult,  
it had problems with the isospin quantum numbers of the lighter fermions,  
and it needed some fine-tuning in order to prevent some fermions from 
acquiring large masses.

 In the present approach, 
 only the mirror particles are coupled strongly and 
 dynamically involved in the 
breaking of $SU(2)_{L}$. By eventually breaking the mirror-generation
symmetries,  small gauge-invariant (by this we mean here and in the 
following gauge-invariant 
under the standard-model gauge group, unless otherwise stated)
masses are allowed 
which communicate the electro-weak symmetry breaking to the 
standard-model fermions by mixing them with their mirror partners. 
This model has neither ``sterile" nor $SU(2)_{L}$-doublet  
light mirror neutrinos, as in \cite{Bag} for example, 
which would pose problems with experiment. 
After the mass hierarchies are computed within this context, phenomenological
consequences like electro-weak precision parameters and  CKM
 matrix elements are then analyzed.  
\section{Matter content}

We start by considering the 
 gauge group structure 
  $SU(4)_{PS} \times SU(2)_{L}\times
SU(3)_{2G} \times U(1)_{G} \times U(1)_{R}$.   
The group $SU(4)_{PS}$ is the usual Pati-Salam group unifying
quarks and leptons, and $SU(2)_{L}$ is the group of weak interactions.
The symmetry  $SU(2)_{R}$ 
has already been broken down to $U(1)_{R}$ by
an  $SU(2)_{R}$-triplet vev at higher 
scales, in order to allow the see-saw mechanism to produce light
Majorana standard-model neutrinos. 

The group $SU(3)_{2G}$ is a horizontal gauge symmetry also
acting only on the mirror
fermions, which becomes strong at around 2 TeV. All other groups are
taken to have weak couplings at this energy. 
The corresponding symmetry for the \newline standard-model fermions  
$SU(3)_{1G}$ 
 has already been broken down to $U(1)_{G}$ at higher scales, at once or 
 sequentially, in order to avoid large FCNC.  

  Under the above gauge structure, 
  the following fermions are introduced, which are left-handed gauge (and not
  mass) eigenstates and transform like 
\begin{displaymath}
\begin{array}{ll} 
{\rm Generations} \\~ \\
\psi_{L}: ( {\bf 4, \;2, \;1,} \;q_{g}, \;0)& \\ 
\psi_{R}^{c}: ( {\bf \bar{4}, \;1, \;1,} \;-q_{g}, \;~^{-1}_{+1} )& \\~\\ 
 {\rm Mirror \;\;\; generations}\\~ \\
 \psi^{M}_{L}:( {\bf 4, \;1, \;3,} \;0, \;~^{+1}_{-1} )& 
\\
 \psi^{M\;c}_{R}: ( {\bf \bar{4}, \;2, \;3,} \;0, \;0)  & 
\end{array}
\nonumber
\end{displaymath}

  \noindent where $g=1,2,3$ is a generation index, with
$q_{1}=\kappa$, $q_{2,3}= -\kappa_{\mp} \equiv
-( \kappa \mp 1)/2$, and $\kappa$ 
an arbitrary abelian charge corresponding to the group $U(1)_{G}$. 
The superscript $M$ denotes the mirror partners
of the ordinary fermions, and $c$ denotes charge conjugation.
  For the sake of compactness  here and in the following 
 the two members of the doublets of the broken $SU(2)_{R}$ symmetry 
 are included in the same parenthesis. 

  One observes that the generation symmetries play a very
  important role at this stage, and this is to  prevent the formation of 
large  gauge-invariant masses. Pairing-up of standard-model and mirror
 generations is thus prohibited, in agreement with what is usually
 called ``survival hypothesis" \cite{Howard}.

 Even though this 
 quantum number assignment is reminiscent of technicolor with
 a strong group 
 $SU(N)_{TC} \approx SU(3)_{2G}$, there is no corresponding
 extended technicolor (ETC) group,
the new anti-particles transform under the same 
(and not the complex conjugate) representation of the
strong group as the new particles, and there is a left-right
 interchange of weak isospin  charges. In addition, the strong
 group in the present case eventually breaks, as it will be
 seen in the following. One should furthermore not  confuse the 
present model with other ``mirror" fermion approaches, like in
\cite{Mex} for example, where all
components of the new fermions are singlets under $SU(2)_{L}$
and interact only gravitationally or marginally with the
standard-model particles, and
which obviously cannot break the electro-weak symmetry dynamically. 

At  high energy  scales that do not enter directly
in this talk, the Pati-Salam group is assumed to break spontaneously like 
$SU(4)_{PS} \times U(1)_{R}  
\longrightarrow SU(3)_{C}\times  U(1)_{Y}$, where $SU(3)_{C}$ and
$U(1)_{Y}$ are the usual QCD and hypercharge groups respectively. 
Much later, 
at scales on the order of $\Lambda_{G} \approx 2$ TeV,   
the mirror generation group breaks sequentially, 
just after it becomes strong, like   
$SU(3)_{2G}\times U(1)_{G} \longrightarrow
SU(2)_{2G}\times U(1)_{G^{\prime}} 
\longrightarrow U(1)^{*}_{G^{\prime \prime}}$, where
the star superscript denotes here and in the following
a broken gauge symmetry.  
We keep track of the $U(1)_{G^{\prime \prime}}$
charges because, even though the corresponding gauge 
group is eventually broken, they could
prove useful to the
qualitative understanding of the fermion mass hierarchies in the model,
as  will be seen later. 
It is not attempted here to investigate
how exactly these breakings occur, and for simplicity it is 
enough to assume that
a Higgs mechanism is responsible for them, effective or not.   
The issue of generation symmetry  breaking
will be discussed again in the discussion section. 
The first spontaneous
generation symmetry breaking $SU(3)_{2G}\times U(1)_{G}
\longrightarrow SU(2)_{2G} \times U(1)_{G^{\prime}}$ occurs at
a scale $\Lambda_{G}$, with an 
$SU(2)_{L}$-singlet scalar state denoted by $\phi_{3}$ 
and transforming like $({\bf 3}, \kappa)$ under 
the generation symmetry acquiring a non-zero vev. 
Note that the group $SU(3)_{2G}$ could in 
principle also self-break dynamically  
{\it via} the fermion-condensation
channel ${\bf 3} \times {\bf 3} \longrightarrow 
{\bf \bar{3}}$ if it were given time to become strongly coupled at this 
energy scale. This breaking channel would however leave the mirror
fermions without $U(1)_{G^{\prime}}$ charge, and we would like to avoid
that for reasons that will become clear shortly.
The fermions have the 
following quantum numbers under the new gauge symmetry 
$SU(3)_{C} \times SU(2)_{L} \times \times SU(2)_{2G}
\times U(1)_{G^{\prime}} \times U(1)_{Y}$:   

\begin{displaymath}
\begin{array}{ll} 
{\rm The \;\;3rd\;\;\&\;\; 2nd \;\;generations} 
\\~\\
q^{3,2}_{L}: ( {\bf 3, \;2, \;1,} \;-\kappa_{\pm}, \;1/3  )& \\~\\ 
l^{3,2}_{L}: ( {\bf 1, \;2, \;1,} \;-\kappa_{\pm}, \;-1  )& \\~\\ 
q^{3,2\;\;c}_{R}: ( {\bf \bar{3}, \;1, \;1,} \;\kappa_{\pm}, 
\;~^{-4/3}_{+2/3} )& \\~\\ 
l^{3,2\;\;c}_{R}: ( {\bf 1, \;1, \;1,} \;\kappa_{\pm}, \;~^{0}_{2} )& 
\\~\\ 
\end{array}
\end{displaymath}

\begin{displaymath}
\begin{array}{ll} 
{\rm The \;\;3rd\;\& \;\;2nd \;\;mirror 
\;\; generations} \\~ \\
 q^{3,2M}_{L}:( {\bf 3, \;1, \;2,} \;-\kappa/2, \;~^{+4/3}_{-2/3} )& 
\\~ \\ 
 l^{3,2M}_{L}: ( {\bf 1, \;1, \;2,} \;-\kappa/2,\;~^{0}_{-2} )& 
\\~ \\ 
 q^{3,2M\;\;c}_{R}: ( {\bf \bar{3}, \;2, \;2,} 
\;\kappa/2, \;-1/3 )&   
\\~ \\
 l^{3,2M\;\;c}_{R}: ( {\bf 1, \;2, \;2,} \;
\kappa/2, \;1 )&  \\~\\~\\ 
\end{array}
\end{displaymath}

\begin{displaymath}
\begin{array}{ll} 
{\rm The \;\;1st \;\;generation\;\;\;\;\;\;\;\;\;\;\;\;} 
\\~\\
q^{1}_{L}: ( {\bf 3, \;2, \;1,} \;\kappa, \;1/3  )& \\~\\ 
l^{1}_{L}: ( {\bf 1, \;2, \;1,} \;\kappa, \;-1  )& \\~\\ 
q^{1\;\;c}_{R}: ( {\bf \bar{3}, \;1, \;1,} \;-\kappa, 
\;~^{-4/3}_{+2/3} )& \\~\\ 
l^{1\;\;c}_{R}: ( {\bf 1, \;1, \;1,} \;-\kappa, \;~^{0}_{2} )& \\~\\ 
{\rm The \;\;1st \;\;mirror 
\;\; generation\;\;\;\;\;\;\;\;\;} \\~ \\
 q^{1M}_{L}:( {\bf 3, \;1, \;1,} \;\kappa, \;~^{+4/3}_{-2/3} )& 
\\~ \\ 
 l^{1M}_{L}: ( {\bf 1, \;1, \;1,} \;\kappa,\;~^{0}_{-2} )& 
\\~ \\ 
 q^{1M\;\;c}_{R}: ( {\bf \bar{3}, \;2, \;1,} \;-\kappa, \;-1/3 )  & 
\\~ \\
 l^{1M\;\;c}_{R}: ( {\bf 1, \;2, \;1,} \;-\kappa, \;1 )&  \\~\\~\\ 
\end{array}
\end{displaymath}

\noindent  where the superscripts 1,...,3 indicate the fermion
generations.
 Moreover, the letters  
$q$ and $l$ stand for 
quarks and leptons respectively.
Note that $\bar{\psi_{R}}\psi^{M}_{L}$ mass
terms are prohibited by the $SU(2)_{2G}$ symmetry for the second and third
generations.

At a scale quite close to 
$\Lambda_{G}$,  the $SU(2)_{2G} \times 
U(1)_{G^{\prime}}$ 
group spontaneously breaks sequentially to $U(1)_{G^{\prime \prime}}$ and this
down to $U(1)^{*}_{G^{\prime \prime}}$ by  
two $SU(2)_{L}$-singlet 
scalar states, denoted by $\phi^{\pm}_{2}$ and  
transforming like $({\bf 2}, \pm 1/2)$ under the generation
symmetry, which  acquire  non-zero vevs.
The quantum numbers of the third and second generation mirror fermions 
after these breakings are given by
\begin{displaymath}
\begin{array}{ll} 
{\rm The \;\;2nd\;\;mirror\;\;generation} \\~\\
 q^{2M}_{L}:( {\bf 3, \;1,} \;-\kappa_{-}, \;~^{+4/3}_{-2/3} )& \\~\\ 
 l^{2M}_{L}: ( {\bf 1, \;1,} \;-\kappa_{-},\;~^{0}_{-2} )& \\~\\ 
 q^{2M\;\;c}_{R}: ( {\bf \bar{3}, \;2,} 
\;\kappa_{-}, \;-1/3 )&  \\~\\ 
 l^{2M\;\;c}_{R}: ( {\bf 1, \;2,} \;
\kappa_{-}, \;1 )& \\~\\ 
{\rm The \;\;3rd \;\;mirror \;\; generation} \\~ \\
 q^{3M}_{L}:( {\bf 3, \;1,} \;-\kappa_{+}, \;~^{+4/3}_{-2/3} )& 
\\~ \\ 
 l^{3M}_{L}: ( {\bf 1, \;1,} \;-\kappa_{+},\;~^{0}_{-2} )& 
\\~ \\ 
 q^{3M\;\;c}_{R}: ( {\bf \bar{3}, \;2,} \;\kappa_{+}, \;-1/3 )  & 
\\~ \\
 l^{3M\;\;c}_{R}: ( {\bf 1, \;2,} \;\kappa_{+}, \;1 )&  \\~\\~\\ 
\end{array}
\end{displaymath}

\noindent while the first mirror generation and all the standard-model 
generation quantum numbers are left unchanged. 
 
The breakings of the mirror generation symmetries   
described above induce at lower energies, among others, effective
four-fermion operators $F$ of the form 
\begin{eqnarray}
F &= & \frac{\lambda}{\Lambda^{2}_{G}}
({\bar \psi^{M}_{R}}\psi^{M}_{L})({\bar \psi^{M}_{L}}\psi^{M}_{R})  
\end{eqnarray}

 \noindent 
 for the three mirror fermion generations, where $\lambda$ 
 are effective four-fermion couplings and the
generation indices are omitted for simplicity. 
The fermion bilinears in parentheses above
 transform like a doublet under $SU(2)_{L}$. 

The next step is to assume that, 
in a manner analogous to top-color scenarios \cite{Hill},    
the $SU(2)_{2G}$ group
is strongly coupled just before it 
breaks,  and it is therefore plausible to take    
the effective four-fermion couplings $\lambda$
 to be critical for the mirror fermions
of the third and second generations, 
 like in the Nambu-Jona-Lasinio model (NJL).    
Therefore, condensates of mirror fermions like 
$<{\bar \psi^{M}_{L}}\psi^{M}_{R}>$   
 can form which  break the symmetry 
$SU(2)_{L}\times U(1)_{Y}$ dynamically down to
the usual $U(1)_{EM}$ group of electromagnetism.

The 
fermion condensates described above  give to the mirror fermions 
symmetry-breaking
masses of order $M  \approx r \Lambda_{F}$  
 {\it via} the operators $F$,   
with $r$  a constant not much smaller than unity if
one wants to avoid excessive fine-tuning of the four-fermion interactions. 
Effective operators of the form 
${\bar \psi^{1\;M}_{R}}\psi^{1\;M}_{L}
{\bar \psi^{2,3\;M}_{L}}\psi^{2,3\;M}_{R}/\Lambda^{2}_{G}$ 
induced by the broken $SU(3)_{2G}$ interaction  feed
down gauge-\newline symmetry-breaking masses to the first 
mirror generation. The fact that all mirror fermions get large masses
of the same order of magnitude 
due to the critical interactions avoids fine-tuning problems that
would appear if mass hierarchies were introduced by allowing only some of 
them to become massive, as is done in \cite{Georgi}. 
Moreover, to avoid breaking QCD and 
electromagnetism, it is assumed that most-attractive-channel
arguments prevent quark-lepton 
condensates of the form \newline $<\bar{q^{M}_{L}}l^{M}_{R}>$
from  appearing. 

If generation symmetries were left intact, the mass matrix ${\cal M}$
for all the
fermions would have the form  
\begin{eqnarray}
&& \;\;\; \psi_{L} \;\;\;\; \psi^{M}_{L} \nonumber \\ 
\begin{array}{c} {\bar \psi}_{R} \\ {\bar \psi^{M}}_{R} \end{array} && 
\left(\begin{array}{cc} 0 & \;\;\;0 \\ 0 & 
\;\;\;M \end{array} \right), 
\nonumber
\end{eqnarray} 

\noindent where the 4 elements shown are blocks of $3\times3$ 
matrices in generation space and $M$ the dynamical mirror-fermion
mass due to the strong generation interactions. 
However, the broken 
generation symmetries allow the formation of gauge-invariant 
masses, and the mass matrix ${\cal M}$ takes the form:
\\~\\ \begin{eqnarray}
&& \;\;\; \psi_{L} \;\;\;\;\;\;\; \psi^{M}_{L} \nonumber \\ 
\begin{array}{c} {\bar \psi}_{R} \\ {\bar \psi^{M}}_{R} \end{array} && 
\left(\begin{array}{cc} 0 & \;\;\;m_{1} \\ m_{2} & 
\;\;\;M \end{array} \right), 
\nonumber
\end{eqnarray}

\noindent  where the diagonal elements  
are gauge-symmetry breaking and the off-diagonal 
gauge-invariant. 

The off-diagonal mass matrices can be generated by
Yukawa couplings $\lambda_{ij}$ associated with spinor
bilinears of fermions with their mirror partners which are
coupled to  the scalar states $\phi_{2,3}$ 
responsible for the  
spontaneous generation symmetry breakings. The corresponding  
gauge-invariant term in the Lagrangian  has the form 
\begin{equation}
\sum_{i,j} \lambda_{ij} \bar{\psi_{iR}}\psi^{M}_{jL} \phi_{2,3},
\end{equation} where
the indices $i,j$ count the corresponding fermions in the model. The
elements of the matrices $m_{1,2}$ will be taken in general to be
quite smaller than
the ones in the matrix $M$, with the exception of the entries
related to the top quark. 

After diagonalization of the mass matrix shown above, in which the lighter
mass eigenstates are identified with the standard-model fermions,
a  {\em see-saw} 
mechanism  produces  small masses for the
ordinary fermions and  larger ones 
for  their mirror partners. A specific 
example for illustration purposes is produced in the next section. 
The situation is reminiscent of universal see-saw models, but 
it involves fermions having quantum-number assignments which should not in
principle pose problems with the Weinberg angle 
$\sin^{2}{\theta}_{W}$ \cite{Bur}, \cite{Cho}.  

Some  remarks relative to the (1,1) block entry of the mass matrix
are in order. First, there are no $<{\bar \psi_{R}}\psi_{L}>$
condensates at these high energy scales. 
Then, after careful inspection
of the quantum numbers carried by the gauge bosons of the broken groups
one observes that there are no four-fermion effective operators
of the form 
$({\bar \psi_{R}}\psi_{L})({\bar \psi^{M}_{L}}\psi^{M}_{R})/\Lambda^{2}_{G}$ 
or any other gauge-invariant operators
for any generation which would feed
gauge-symmetry-breaking  
masses to the ordinary fermions.
\section{Hierarchies, mixings and precision tests} 

The mass hierarchies produced by the model are computed next, since
they provide the basis of any phenomenological analysis. 
The gauge-symmetry breaking mass submatrices $M$ 
are hermitian because of parity symmetry.
 The gauge-invariant ones, denoted by $m_{1,2}$ should be symmetric
 due to the quantum numbers assigned to the fermions,   
 but not necessarily real. 
Complex matrix elements allow therefore
in general  for weak CP violation.
Assuming that $SU(2)_{L}$ effects can be  
neglected in the gauge-invariant mass generation process or
that their effect is just homogeneously multiplicative, one also has 
the relation $m_{2}= c\,m^{\dagger}_{1}$ between the
gauge-invariant submatrices, with $c$ a real constant. 
This  means that the determinant of the mass matrix ${\cal M}$ is real, 
eliminating thus the strong CP problem in this approximation,
at least at tree level.

For simplicity, the mass matrices in the following are  taken real and
having the form 
\begin{eqnarray}
&& {\cal M}_{i} = \left(\begin{array}{cc}  0 & m_{i} \\ m_{i} & M_{i} 
 \end{array} \right), i = U,D,l    
\end{eqnarray}

\noindent for the up-type quarks ($U$), down-type quarks ($D$) and charged
leptons ($l$). We give as a numerical example 
forms for the off-diagonal gauge-invariant 
mass submatrices of the up-type and down-type quark
sectors for illustration purposes (with obvious correspondence
between column and row numbers with generation indices):   
\begin{eqnarray}
&& \begin{array}{c}  \\ m_{U} ({\rm GeV}) =\\ \end{array}
\left(\begin{array}{ccc} 2.3 & 5.7 & 1.1 \\ 5.7 & 20 & 1.3 \\
1.1 & 1.3 & 360 \end{array} \right)  \nonumber \\ 
&& \begin{array}{c}  \\  m_{D} ({\rm GeV}) =\\ \end{array} 
\left(\begin{array}{ccc} 1.6 & 1.6 & 0.51 \\ 1.6 & 4 & 1.3 \\
0.51 & 1.3 & 35 \end{array} \right).    
\end{eqnarray}

\noindent Without loss of generality, 
the dynamical assumption is made here that the $SU(2)_{L}$-breaking 
mass submatrices are diagonal and have the form 
\begin{eqnarray}
&& \begin{array}{c}  \\ M_{U} ({\rm GeV}) =\\ \end{array} 
\left(\begin{array}{ccc} 360 & 0 & 0 \\ 0 & 650 & 0 \\
0 & 0 & 650 \end{array} \right) \nonumber \\ 
&& \begin{array}{c}  \\  M_{D} ({\rm GeV}) =\\ \end{array} 
\left(\begin{array}{ccc} 200 & 0 & 0 \\ 0 & 360 & 0 \\
0 & 0 & 360 \end{array} \right).  
\end{eqnarray}

\noindent The gauge-symmetry breaking masses of the first mirror generation
are taken to be smaller than the ones of the two heavier
generations because they are fed down by effective operators
that are not critical like the ones for the other mirror generations. 

 It is  also expected that the dynamics provide some 
 custodial symmetry breaking which is responsible for 
 the mass difference in the up- and down-quark sectors. The $U(1)_{Y}$ 
 could be in principle the source of this difference, but 
 we do not speculate on how this is precisely realised here. 
One has to further stress that the splitting of $M_{U}$ and
$M_{D}$ is not {\it a priori} needed to produce the
top-bottom quark mass hierarchy , but it is introduced only to
better fit the experimental constrains on the 
electro-weak parameters, as  will be seen later.

These mass matrices give, after diagonalization and without the need for
any fine-tuning,  the following
quark and mirror-quark masses  (given in units of GeV):  
 
\noindent 
Standard-model quarks \hfill Mirror quarks $\;$ \\ 
$m_{t} = 160$, $m_{c} = 0.77$, $m_{u} = 0.001$  \hfill 
$m_{t^{M}} = 810$, $m_{c^{M}} = 651$, $m_{u^{M}} = 360\;$  \\
$m_{b} = 3.4\;$, $m_{s} = 0.07$, $m_{d} = 0.003$  \hfill 
$m_{b^{M}} = 363$, $m_{s^{M}} = 360$, $m_{d^{M}} = 200$. 
\newline
The ordinary quark masses given are slightly smaller than
the ones usually quoted because the values reported here are relevant
to the characteristic scale of the new strong dynamics which is around 
2 TeV, and one has therefore to account for their running with
energy. The formalism 
presents no inherent difficulty whatsoever producing larger masses
for these fermions. 

The generalization of the standard-model CKM quark-mixing matrix in this 
scenario is a unitary 
$6\times 6$ matrix of the form $V_{G}=K^{T}_{U}K_{D}$, where the matrices
$K_{U,D}$ diagonalize the fermion mass matrices like 
${\cal M}_{i}=K^{T}_{i}J_{i}K_{i}$, $i=U,D$,  with $J_{U,D}$ 
being the two $6 \times 6$ diagonal mass matrices of the up- and down-quark
sector.  
The generalized 
CKM matrix has the form 
\begin{eqnarray}
 V_{G}= \left(\begin{array}{cc} V_{CKM} & \;\;\; V_{1} \\ V_{2} & 
\;\;\; V^{M}_{CKM} \end{array} \right),  
\end{eqnarray} 

\noindent and the usual standard-model CKM matrix $V_{CKM}$ is one of its 
submatrices given (in absolute values) by 
\begin{eqnarray}
\begin{array}{c}  \\ | V_{CKM} | =\\ \end{array} &&
\left(\begin{array}{ccc} 0.98 & 0.22 & 0.003 \\ 0.22 & 0.97 & 0.042 \\
0.006 & 0.038 & 0.95 \end{array} \right),   
\end{eqnarray}

\noindent which is consistent with present experimental
constraints. 

The mixing between the first and second generations is larger
than the one between the second and third generations, and this
can be easily traced back to the relative elements of $m_{U,D}$. 
Furthermore, one has to be particularly cautious 
when using the flavor symbol `t' and the flavor name `top quark'
for the heaviest standard-model-quark mass eigenstate, since  
$t_{L} (t^{c}_{R})$ has a non-negligible $SU(2)_{L}$ 
singlet (doublet) component as expected due to the large
$\bar{t_{R}}t^{M}_{L}=\bar{t^{M}_{R}}t_{L}$ mass terms, and this 
is reflected on the reported value
of $|V_{tb}|=0.95$. This is particularly apparent in the
third-generation fermions to which correspond larger gauge-invariant masses, 
since the fermion-mirror fermion 
mixings are given  roughly by the ratio $m_{ii}/M_{ii}$.  
Present experimental data give $|V_{tb}|=0.99 \pm 0.15$ \cite{Tarta}. 
More precise future measurements of
this quantity should show deviations from its standard-model value which
is very close to 1 assuming unitarity of the mixing matrix $V_{CKM}$.  
Larger mirror-fermion masses can diminish this effect by
reducing the corresponding mixing of the mirrors with the
ordinary fermions. 

In fact, indirect experimental indications for the
existence of $SU(2)_{L}$-singlet
new fermions which can mix with the third standard-model-generation 
charged fermions
$t_{L}$, $b_{L}$, $\tau_{L}$, and $SU(2)_{L}$-doublet 
new anti-fermions which can mix 
with $t^{c}_{R}$, $b^{c}_{R}$ and $\tau^{c}_{R}$ 
could already exist in LEP/SLC precision data. 
One would be coming from the $S$ and $T$ parameters, which are 
consistent with anomalous top-quark couplings, as will be seen later, 
and the other coming from anomalous $b$-quark and $\tau$-lepton
couplings  to the $Z^{0}$ boson corresponding to even $3 \sigma$ effects 
\cite{Field}. 
The actual sign of the deviations depends on the relevant interaction 
strength of the two isospin partners of the mirror doublets with
the standard-model fermions, but more details
on  this are given later. Deviations from  the weak couplings 
of the lighter standard-model particles are heavily suppressed,  
but they can be potentially large when the mirror partners are light. 
Bringing all the mirror partners to lower scales 
should be avoided nevertheless, since reproducing the weak scale
would then require fine-tuning, as will be shown shortly.

The corresponding CKM matrix for the 
mirror sector $V^{M}_{CKM}$ is equal (in absolute values) to 
\begin{eqnarray}
\begin{array}{c}  \\ | V^{M}_{CKM} | =\\ \end{array} &&
\left(\begin{array}{ccc} 1 & 0.001 & 0.001 \\ 0.001 & 1 & 0.039 \\
0.001 & 0.036 & 0.95 \end{array} \right).    
\end{eqnarray}

\noindent 
The third generation is here the main reason  why this matrix is not diagonal
(The entries (1,1) and (2,2) are close to unity because
of the assumed diagonal form of $M_{U,D}$, but not exactly unity, so  
that the unitarity character of the mixing matrix $V_{G}$ is preserved.) 
Furthermore, the matrices 
$V_{1}$ and $V_{2}$ mix the up-quark sector of the 
standard model with the down mirror-quark 
sector and {\it vice-versa}, but most of their entries are quite small
and we do not list them here.

For the charged leptons,  a diagonal gauge-symmetry 
breaking mass matrix  is used again and  a
gauge-invariant mass matrix having the  forms    
\begin{eqnarray}
&& \begin{array}{c}  \\ M_{l} ({\rm GeV}) =\\ \end{array} 
\left(\begin{array}{ccc} 180 & 0 & 0 \\ 0 & 200 & 0 \\
0 & 0 & 200 \end{array} \right) \nonumber \\ 
&& \begin{array}{c}  \\ m_{l} ({\rm GeV}) =\\ \end{array}
\left(\begin{array}{ccc} 0.25 & 0.25 & 0.1 \\ 0.25 & 3.8 & 1 \\
0.1 & 1 & 17 \end{array} \right).  
\end{eqnarray}

\noindent These give the following lepton and mirror-lepton 
mass hierarchy (at 2 TeV and in GeV units): \\  
Standard-model charged leptons \hfill Mirror charged leptons $\;$ \\ 
$m_{\tau} =1.45$ , $m_{\mu}=0.07$,   
$m_{e}=3\times10^{-4}$  \hfill   
$m_{\tau^{M}} = 201$,  
$m_{\mu^{M}} = 200$, $m_{e^{M}} = 180$. \newline 
The difference of the charged-lepton mass matrix with the
down-quark mass matrix is attributed to QCD effects.
The same mass hierarchies could have been produced with
a diagonal submatrix $m_{l}$ which would require less parameters, 
but for the sake of consistency a submatrix form similar to $m_{D}$
is chosen.
The neutrino mass and mixing
matrix could be quite interesting and
is left for future work, since the fact that neutrinos 
can have both Dirac and Majorana masses makes theoretical 
considerations and calculations more involved.

We next proceed by giving an estimate
for the dynamically generated weak scale $v$. 
A rough calculation using the Pagels-Stokar formula gives 
\begin{equation}
v^{2} \approx \frac{1}{4\pi^{2}} 
\sum^{N}_{i} M_{i}^{2} \ln{(\Lambda_{G}/M_{i})} \;, 
\end{equation}

\noindent where $N$ is the number of new weak doublets introduced and
$M_{i}$ their mass, where it has been  assumed for simplicity that
$m_{\nu^{M}_{i}} = m_{u^{M}}$ for all mirror neutrinos and where
departures from pure weak eigenstates have been neglected. 
Consequently, 
for the masses found before and  $\Lambda_{G} \approx $ 1.8 TeV
 one gets $v \approx 250$ GeV, as is required.  
The mirror fermions can therefore be heavy enough to eliminate any
need for excessive fine-tuning of the four-fermion interactions responsible
for their masses.  

The $S$ parameter \cite{Tatsu} could be problematic in this scenario
however, since 
12 new $SU(2)_{L}$ doublets are introduced. 
 The main negative effect able to cancel the corresponding
 large positive contributions to $S$ coming from ``oblique" corrections 
 is the existence of vertex corrections 
 stemming from 4-fermion
effective interactions, which can give rise to  similar effects as   
the ones induced by  light $SU(2)_{L}$-invariant scalars 
known as ``techniscalars" \cite{Kagan}.    

More precisely, it is argued that the effective Lagrangian of the theory,  
after the spontaneous breaking of the $U(1)_{G^{\prime \prime}}$  
generation symmetry, 
contains terms which can lead to a shift to the couplings of the top and 
bottom quarks to the $W^{\pm}$ and $Z^{0}$ bosons.   
In particular, there are four-fermion terms
involving 3rd generation-quark flavor eigenstates and their
mirror partners given by  
\begin{eqnarray}
&&{\cal L}_{{\rm eff}} = - \left(
\frac{\lambda_{n1}}{\Lambda^{2}_{n1}} 
\bar{t^{M}_{L}}\gamma^{\mu}t^{M}_{L}
+ \frac{\lambda_{c1}}{\Lambda^{2}_{c1}}
\bar{b^{M}_{L}}\gamma^{\mu}b^{M}_{L}
\right)
\bar{t_{R}}\gamma_{\mu}t_{R} \nonumber - \\   
&&-\left( 
\frac{\lambda_{c2}}{\Lambda^{2}_{c2}} 
\bar{t^{M}_{L}}\gamma^{\mu}t^{M}_{L}
+ \frac{\lambda_{n2}}{\Lambda^{2}_{n2}} 
\bar{b^{M}_{L}}\gamma^{\mu}b^{M}_{L}
\right) \bar{b_{R}}\gamma_{\mu}b_{R}
\end{eqnarray}

\noindent where the $\lambda$'s and
$\Lambda$'s are
the effective positive couplings and scales of the 
corresponding operators renormalized at the $Z^{0}$ boson mass, and
the subscripts $n,c$ indicate whether the participating fermions have
the same hypercharge or not. Note that terms like  
$\frac{\lambda_{n3}}{\Lambda^{2}_{n3}}  
(\bar{q^{M}_{R}}\tau^{a}\gamma^{\mu}q^{M}_{R}) 
(\bar{q_{L}}\tau^{a}\gamma_{\mu}q_{L})$,  
where  $\tau^{a}, a=1,2,3$ are the three $SU(2)_{L}$ generators, 
cannot be generated here in perturbation theory,
unlike analogous terms in extended technicolor models. Anyway,
such terms would 
produce shifts only to the left-handed fermion couplings, and these 
are already too much constrained from LEP/SLC 
data to be of any interest here. 

Adopting the effective Lagrangian approach for the 
heavy, strongly interacting sector of the theory \cite{Sekhar},  
the two mirror-fermion currents are expressed in terms of effective 
chiral fields $\Sigma$ like  
\begin{eqnarray}
 \bar{t^{M}_{L}}\gamma^{\mu}t^{M}_{L}  
&=& i \frac{v^{2}}{2}{\rm Tr}\left(\Sigma^{\dagger}
\frac{1+\tau^{3}}{2}D^{\mu}\Sigma \right) 
 \nonumber \\ 
 \bar{b^{M}_{L}}\gamma^{\mu}b^{M}_{L} 
&=& i \frac{v^{2}}{2}{\rm Tr}\left(\Sigma^{\dagger}
\frac{1-\tau^{3}}{2}D^{\mu}\Sigma \right) 
\end{eqnarray}

\noindent where the covariant derivative $D^{\mu}$ is defined by  
\begin{equation}
D^{\mu}\Sigma = \partial^{\mu}\Sigma +ig \frac{\tau^{a}}{2}W_{a}^{\mu}\Sigma 
-ig^{\prime} \Sigma \frac{\tau^{3}}{2}B^{\mu}.   
\end{equation}

\noindent 
The $g$ and $g^{\prime}$ above are the couplings
corresponding to the gauge fields $W^{\mu}_{a}$ and $B^{\mu}$ of
the groups $SU(2)_{L}$ and $U(1)_{Y}$ respectively.
The chiral field 
$\Sigma=e^{2i \tilde{\pi}/v}$ transforms like $L\Sigma R^{\dagger}$
with $L \in SU(2)_{L}$ and $R \in  U(1)_{Y}$ as usual, 
with hypercharge $Y=\tau^{3}/2$ and
$\tilde{\pi}= \tau^{a}\pi^{a}/2$ containing the would-be
Nambu-Goldstone modes $\pi^{a}$ ``eaten" by the electro-weak bosons. 

In the unitary gauge $\Sigma=1$, and the currents given 
above induce shifts in the standard-model Lagrangian
of the form
\begin{eqnarray}
\delta {\cal L}&=&(gW^{\mu}_{3}-g^{\prime}
B^{\mu})\left ( \delta g^{t}_{R}
\bar{t_{R}}\gamma_{\mu}t_{R} + \right. \nonumber \\ 
&& \left. \delta g^{b}_{R}\bar{b_{R}}\gamma_{\mu}b_{R} \right) 
\end{eqnarray} 

\noindent 
with the non-standard fermion-gauge boson couplings expressed by  
\begin{eqnarray}
\delta g^{t}_{R} &=& \;\; \frac{v^{2}}{4}
\left(\frac{\lambda_{n1}}{\Lambda^{2}_{n1}}-
\frac{\lambda_{c1}}{\Lambda^{2}_{c1}}\right) \nonumber \\ 
\delta g^{b}_{R} &=& 
- \frac{v^{2}}{4}
\left(
\frac{\lambda_{n2}}{\Lambda^{2}_{n2}} 
-\frac{\lambda_{c2}}{\Lambda^{2}_{c2}} \right). 
\end{eqnarray}

\noindent 

After Fierz rearrangement of the terms in the 
effective Lagrangian ${\cal L}_{{\rm eff}}$,  
the scales $\Lambda_{n1,n2,c1,c2}$ can be seen as masses of 
effective scalar $SU(2)_{L}$-singlet
spinor bilinears consisting of a 
mirror and an ordinary fermion. These are
reminiscent of ``techniscalars" as to their quantum numbers.  
The effective four-fermion couplings are not
only determined by
 the corresponding $U(1)_{G}$  charges,
since the present situation is closer to gauged NJL models where
unbroken strong gauge interactions can influence considerably four-fermion
terms.  One may observe that, unlike the 
present situation, the corresponding scalar effective
operators induced by ETC
interactions in ordinary technicolor theories would be
$SU(2)_{L}$-doublets and would not produce shifts to the 
right-handed fermion couplings. The scalar operators appearing here
could in principle correspond to mesons bound with the
QCD force, but their constituents are very heavy and 
are expected in principle to decay weakly before they have time to 
hadronize.

It is as a matter of fact difficult to predict the values of
the effective couplings of the operators that determine the  
fermion anomalous couplings, since 
they are influenced by non-perturbative dynamics.   
The values of the various terms  
are  here chosen for illustration purposes to be  
$\frac{\lambda_{n1}v^{2}}{\Lambda^{2}_{n1}}=1$,   
$\frac{\lambda_{c1}v^{2}}{\Lambda^{2}_{c1}}=3.32$,   
$\frac{\lambda_{n2}v^{2}}{\Lambda^{2}_{n2}}=0.22$,   
$\frac{\lambda_{c2}v^{2}}{\Lambda^{2}_{c2}}=0.1$.  
The terms corresponding to operators involving the standard-model top
quark (see subscripts $n1, c1$)
are assumed larger than the ones involving the standard-model bottom 
quark (subscripts $n2, c2$). 
This might be related to the fact that, as was already seen 
in the mass matrices, $\bar{t_{R}}t^{M}_{L}$ gauge-invariant
mass terms are much larger than 
$\bar{b_{R}}b^{M}_{L}$ terms, which is needed in order to reproduce the
correct top-bottom mass hierarchy. 
This would also explain why four-fermion terms involving first- and
second-generation quarks are neglected in this analysis. 

Using the values above  one finds the anomalous couplings 
 $\delta g^{b}_{R}=-0.03$ and  $\delta g^{t}_{R}=-0.58$.  
The coupling $\delta g^{b}_{R}$ 
is  within  its best-fit 
value $\delta g^{b}_{R}  = 0.036 \pm 0.068$ 
(this is a combined fit including
information on $\delta g_{L}$ and the $S$ and $T$ parameters 
\cite{DoTe}). It works here against $\delta g^{t}_{R}$ since it 
contributes positively, by a comparatively small amount,
to the $S$ parameter. It is already so tightly constrained 
that, even if it finally turns out to be positive, as suggested by
\cite{Field} and which is 
easily achievable here by an appropriate choice of the
relevant four-fermion couplings, it
will not change our conclusions substantially. 
The coupling $\delta g^{t}_{R}$ is of course not yet constrained, and
it is therefore a good candidate for 
a possible source of the large vertex corrections needed in this model.  

One should expect therefore
that, apart from the
model-independent ``oblique" contributions to the electro-weak 
precision parameters $S$ and $T = \Delta\rho / \alpha$ (where
$\alpha$ is the fine structure constant), denoted by 
$S^{0}$ and $T^{0}$, these parameters   
receive also important vertex corrections 
$S^{t,b}$ and $T^{t,b}$ due to the top and bottom quarks, 
which should be  given in terms of the anomalous couplings 
calculated above. 
The ``oblique" positive corrections to $S$ are given by $S^{0}=0.1N$
for $N$ new $SU(2)_{L}$ doublets, assuming QCD-like strong dynamics. 
On the other hand, the mass
difference  between the up- and down-type mirror fermions 
produces a positive contribution to $T^{0}$. Considerations in the past
literature with mirror
fermions or vector-like models 
which can give very small or negative $S^{0}$ and $T^{0}$ do not 
concern us here because they are, unlike the present case, based on
the decoupling theorem due to the existence of
large gauge-invariant masses \cite{litri}, \cite{many}. 

By summing up these effects therefore, one finds for $S$ and $T$ 
the expressions \cite{DoTe}  
\begin{eqnarray}
S&=&S^{0}+S^{t,b} = 0.1N 
+ \nonumber \\ && \frac{4}{3\pi}(2 \delta g^{t}_{R} - \delta g^{b}_{R})   
\ln{(\Lambda/M_{Z})} \nonumber \\   
 T&=&T^{0}+T^{t,b} = 
\frac{3}{16\pi^{2}\alpha v^{2}}\sum_{i}^{N}(m_{U_{i}^{M}}-m_{D_{i}^{M}})^{2}
+\nonumber \\ && \delta g^{t}_{R}
\frac{3m^{2}_{t}}{\pi^{2}\alpha v^{2}} \ln{(\Lambda/m_{t})}, 
\end{eqnarray}

\noindent where $m_{U_{i}^{M},D_{i}^{M}}$ denote the masses of the
up- and down-type mirror
quarks, $N=12$ in the present case, 
and $\Lambda$ is the cut-off,   
which is  expected to be the smallest of the scales
$\Lambda_{n1,n2,c1,c2}$. Note that
these expressions are valid for small anomalous couplings, but they
are used in the following to illustrate the main effect of new sector
even though the top-quark anomalous coupling is taken to be quite large.
Moreover, there should be corrections to these
formulas, mainly related to the
mirror top quark, due to the fact that the
mirror mass eigenstates are not pure gauge eigenstates. They are 
in the following neglected since they should be -at least partially-
compensated by the fact that the top quark is also not a weak 
eigenstate but has a weak-singlet admixture.  
It is also noted that 
contributions to $S^{0}$ and $T^{0}$ from the lepton sector
are calculated assuming Dirac mirror neutrinos. 

Nevertheless, one has to stress here that no isospin
splitting whatsoever is required {\it a priori}
in the mirror sector in order to get the
top-bottom quark mass hierarchy, since this can be produced by
differences in the gauge-invariant mass submatrices. 
The dynamical generation of this hierarchy  does not lead
to problems with the $T$ parameter, and this can be traced to the  fact that
the fermion 
condensates which break dynamically the electro-weak symmetry are  
distinct from the electro-weak-singlet
condensates responsible for the feeding-down of 
masses to the standard-model fermions. This is contrary to the
usual ETC philosophy and closer to the conceptual basis of \cite{Bob}. 
The reason 
this isospin asymmetry is introduced here is only to cancel the large negative
contributions to the $T$ parameter coming from the vertex corrections, 
as  will be seen in the following. 

By using the fermion masses and anomalous couplings  
calculated above, one finds that the parameters $S$ and $T$ are given by 
\begin{eqnarray}
S & \approx & 1.2 -0.48 \ln{(\Lambda/M_{Z})}  
\nonumber \\  
T & \approx & 19.4 \times ( 0.88 - 0.58 \ln{(\Lambda/m_{t})}). 
\end{eqnarray} 

\noindent 
The present best-fit  values for the  electroweak parameters  are 
(note that this is again a
combined fit including b-quark anomalous-coupling information
\cite{DoTe})  
\begin{eqnarray}
S & = & -0.40 \pm 0.55 \nonumber \\
T & = & -0.25 \pm 0.46 \;. 
\end{eqnarray} 

\noindent 
One observes therefore that for cut-off scales $\Lambda$ equal or
larger than
about $ \Lambda \approx 0.8$ TeV, which is the smallest of the scales 
$\Lambda_{n1,n2,c1,c2}$, 
negative values for the $S$ and $T$ parameters consistent 
with experiment are
feasible, i.e. 
$S ~^{<}_{\sim}\; 0.14$, $T ~^{<}_{\sim}\; -0.3$, 
and this is mainly due to the large negative anomalous coupling
$\delta g^{t}_{R}$.
Similar values for the electro-weak precision parameters
could be achieved with smaller anomalous couplings accompanied with
a larger cut-off $\Lambda$. This would lead to lighter 
mirror fermions in order to reproduce the weak scale correctly,
something that would also automatically imply a larger fermion-mirror fermion
mixing,  
but it would have the undesirable effect of increasing the fine tuning
in the model.

One way to achieve a smaller $S$ parameter is to note that
the generation group is broken, leading to non-QCD-like strong dynamics.
If this makes the mirror-fermion masses run much slower with 
momentum, it can reduce the positive 
contributions to the $S$ parameter even by a factor of two \cite{ApGe}.
In any case, the purpose of the numerical example presented 
is merely to illustrate that theories of this type may 
potentially  produce negative $S$ and $T$ parameters. 
\section{Discussion}

A possible  origin of the gauge group
structure introduced in this study and  a breaking mechanism of 
the mirror generation groups are  discussed next.
The following unification gauge group is considered  
 \begin{displaymath}
\hspace{-0.4in}
SO(10)_{1}\times SU(4)_{1G} \newline \\ \times SO(10)_{2} \times SU(4)_{2G} 
\subset E_{8_{1}} \times E_{8_{2}}, 
\end{displaymath}  under which the matter fields,
contained initially in the adjoint representation of the $E_{8}$
groups, transform like ${\bf (16, \;\bar{4}, \;1, \;1)}$ 
and ${\bf (1, \;1, \;\bar{16}, \;4)}$ for the ordinary and mirror 
fermions respectively.  This corresponds to four ordinary and
four mirror-fermion generations including $SU(2)_{L}$-singlet neutrinos. 
Such a fermion content is Witten-anomaly free \cite{Witten}.
One has then  the subsequent spontaneous breaking
of   $SO(10)_{i} \times SU(4)_{iG}
\longrightarrow G_{i}  \times SU(3)_{iG}$ 
at unification scales, at once or sequentially,  
where $G_{i} \sim SU(4)_{iPS} \times  SU(2)_{iL} \times SU(2)_{iR}
\times U(1)_{iG}$, 
with $i = 1, 2$.    
In this context the three lighter mirror-fermion
generations could have $U(1)_{2G}$ couplings from the start, with
a natural value for the $U(1)_{iG}$ charges being $\kappa = 1/3$. 
 Still at unification scales, the group  
 $G_{1} \times G_{2}$  breaks spontaneously down to its diagonal
subgroup  $G_{D} \sim SU(4)_{PS} 
\times SU(2)_{L} \times SU(2)_{R}\times U(1)_{G}$.   
Subsequently, the fourth generations can pair-up and disappear from 
the low energy spectrum.

In addition, the $SU(2)_{R}$ group is assumed 
to break spontaneously 
down to  $U(1)_{R}$ {\it via} an $SU(2)_{R}$ triplet, in order 
to allow for a see-saw mechanism for the standard-model neutrino masses. 
The gauged generation symmetry 
$SU(3)_{1G}$ acting on the standard-model
fermions has also to be broken spontaneously 
at high energy scales, at once or sequentially,
in order to avoid large direct FCNC in the standard-model sector and
to prevent the pairing-up of the fermion generations with their
mirror partners. 
These two breakings signal parity violation, 
and are at the source of the fundamental
asymmetry between ordinary and mirror fermions in nature. 
The unbroken $SU(3)_{2G}$ will allow the condensation of 
mirror fermions at  much lower energy scales, and is 
remotely reminiscent of the ``heavy color" group that was considered in 
\cite{Franck} in order to conceal these new fermions. The
difference here is that, far from concealing the mirror 
partners, we mix them with the ordinary fermions by eventually
breaking the generation group, in order to generate masses for
the standard-model particles. One can further note here that
the gauge sector responsible for the eventual mirror-fermion
condensation at lower energy scales
corresponds to what is usually called  ``hidden sector" used
for gaugino condensation in dynamical supersymmetry breaking models. 

Since the breaking of the generation groups takes
place just after the corresponding gauge couplings become strong, 
we speculate next on how a dynamical mechanism 
could be responsible for this effect. 
It is  namely observed that operators of the form 
$\bar{\psi^{M}_{L}}\psi_{R}$ 
 transform like 
a $({\bf 3}, \kappa)$ under $SU(3)_{2G}\times U(1)_{G}$ 
and are singlets under the standard-model gauge symmetry.
They have therefore the same quantum numbers as the scalar
state $\phi_{3}$  introduced in this talk. 
If these composite operators of fermions could gain non-zero vevs  
they would  break the mirror generation symmetry dynamically to 
$SU(2)_{2G} \times U(1)_{G^{\prime}}$. 

Operators of the same form,
transforming like $({\bf 2}, \pm 1/2)$ under the new generation 
symmetry and having the same quantum numbers as $\phi^{\pm}_{2}$
would  break dynamically this gauge symmetry completely if they could also  
acquire subsequently non-zero vevs. 
(Note that if $U(1)_{G^{\prime \prime}}$ breaking occurs dynamically in
the same fashion
one would have to introduce for consistency 
slightly non-diagonal gauge-symmetry
breaking mass submatrices $M_{U,D}$, which would however not alter
qualitatively the results reported. Large non-diagonal 
elements both in the invariant and symmetry-breaking submatrices should
be avoided nevertheless, since they would 
create problems with the quasi-diagonal $V_{CKM}$.)
It would be therefore possible to identify these fermionic
condensates 
with the scalars introduced like \newline  $<\phi_{2,3}> \approx \frac{
<\bar{\psi^{M}_{L}}\psi_{R}>}{\Lambda^{2}_{G}}$. 

The only strong interaction 
able to generate such condensates in the present framework
is unfortunately QCD. It could be
of course assisted by non-negligible $U(1)_{G}$
interactions. The problem is that a strongly
coupled abelian group at the TeV scale
could in principle pose problems with a Landau pole
below the unification scale, unless it is soon embedded in a
non-abelian group
or other  unknown dynamics are involved in the process.  
On the other hand, 
non-abelian instead of abelian
generation groups common to ordinary and mirror
fermions are also problematic. A common $SU(2)_{G}$  would not
have a negative $\beta$-function, and a common $SU(3)_{G}$ would
pose problems with too large FCNC in the first generation.
The question is therefore  whether  
QCD could be responsible for the generation symmetry breaking.   
One would then
have a situation where the generation symmetry breaks due to 
strong dynamics at scales much smaller 
than the scale $\Lambda_{G}$ where it becomes strong, 
since the QCD characteristic scale $\Lambda_{QCD}$ 
is on the order of 1 GeV,
which is similar to the situation discussed in \cite{litri}.
Examples for scenarios of this kind, even though {\it a priori} not 
excluded, have not been seen in nature yet.  

However, in order to make such a scheme
consistent with what has been already discussed,  
non-perturbative
effects should shift the masses of the bosons corresponding to 
the broken symmetries of the generation group from $\Lambda_{QCD}$ 
roughly
three orders of magnitude up to its characteristic scale $\Lambda_{G}$. 
It is the fact that the generation
group has already become strongly coupled at higher scales
that allows us to speculate
that the condensate $<\bar{\psi^{M}_{L}}\psi_{R}>$, 
 if it ever forms
due to QCD and to $U(1)_{G}$, and possibly having a large top-quark component,
could acquire values much larger than the QCD scale by such effects. 
An argument supporting this view could be that,  
 after the generation group breaking, there is no symmetry 
protecting   gauge-invariant composite operators 
from  acquiring large vevs, and radiative corrections
could in principle shift these vevs anywhere from $\Lambda_{QCD}$
to $\Lambda_{G}$. 

This hierarchy of scales
between $\Lambda_{QCD}$ and $\Lambda_{G}$ would also 
potentially explain without fine-tuning
the smallness of most of the gauge-invariant masses in comparison 
with the gauge-symmetry breaking ones, with 
the exception of the top and bottom quarks. 
In a scenario of this type leptons
would get their masses {\it via } four-fermion operators induced by the
broken Pati-Salam group.
In any case, if QCD, possibly assisted by $U(1)_{G}$, is unrelated to the  
formation of  such condensates because of the large energy-scales discrepancy, 
the only way to avoid fundamental Higgs
fields in this approach would be a new strong interaction not present in 
this scenario or other unknown dynamics. 

To summarize, we were motivated by several theoretical arguments 
and possibly by some experimental  
indications that there are new physics around
the TeV scale, and we showed how one can extend 
the gauge sector of the standard model
and its fermionic content in a left-right symmetric context. We argue 
that doubling  the matter degrees of freedom should be considered 
positively if, instead of just burdening the theory with more parameters,
it renders it more symmetric while simultaneously solving several problems
like fine-tuning, mass generation, and possibly 
absence of strong CP violation and eventual unification. 

It was shown that the model    
  sets up a precise theoretical framework for the calculation of
fermion mass hierarchies and mixings.
It gives furthermore rise to dynamics which could 
potentially reconcile the $S$- and $T$- parameter
theoretical estimates  with their experimental values without
excessive fine tuning.  Moreover, the doubling  
of the fermionic spectrum it predicts provides decay modes which
should in principle be detectable  
in colliders like $LHC$ and $NLC$. This fact, together
with more precise future 
measurements of possible FCNC and anomalous couplings in 
the third fermion generation render the model  experimentally
testable.

Within the present approach, a deeper understanding of the generation of  
the gauge-invariant mass matrices $m$ and the
effective couplings leading to anomalous third-generation
standard-model fermion couplings to the $Z^{0}$ boson is still
needed. This would settle the question on whether the large positive
loop corrections to the $S$ parameter in this model can be adequately
canceled by vertex corrections without unnatural fine-tuning.
Furthermore, the investigation on how  
the mirror generation groups break just after they
become strong and  the
unification of couplings at high energies in a way
consistent  with present bounds on the proton life-time are 
important questions left for future studies. 
\acknowledgments{ 
I thank S. Chivukula, M. Lindner, N. Maekawa and V. Miransky
for very helpful discussions. 
The speaker is supported by an {\it Alexander von Humboldt Fellowship}.}

\end{document}